# A laboratory investigation of thermally induced pore pressures in the Callovo-Oxfordian Claystone

M. Mohajerani[1,], P. Delage[1], J. Sulem[1], M. Monfared[1], A.M. Tang[1], B. Gatmiri[2]

1. Ecole des Ponts ParisTech, UR Navier/CERMES, 6-8 av. B. Pascal, F 77455 Marne la Vallée cdx 2
2. ANDRA, Châtenay Malabry

## Abstract

In the framework of research into radioactive waste disposal, it was decided to investigate the thermally induce pore pressure occurring in the Callovo-Oxfordian claystone, a possible host rock in which the ANDRA underground laboratory of Bure (East of France) has been excavated. Thermal pore pressures appear in low permeability soils and rocks because the thermal expansion coefficient of water is significantly higher than that of the solid grains (Campanella and Mitchell; 1968 [1], Ghabezloo and Sulem; 2009 [2]). This phenomenon has clearly been observed in various in-situ heating tests conducted in Opalinus claystone in the Mont-Terri Underground Research Laboratory (URL) in Switzerland (HE-D test) and in Callovo-Oxfordian (COx) claystone in the Bure URL in France (TER test, Wileveau and Su; 2007 [3])

The processes of coring, transportation, storage and specimen trimming induce some desaturation in the sample. Due to the very low permeability ($10^{-20}$ m$^2$) of the COx claystone, a long period of time is necessary to properly resaturate the sample, a mandatory condition to satisfactorily investigate thermal pressurisation. Particular emphasis was hence put on the previous saturation procedure that was carried out under in-situ effective stress condition.

Thermal pressurization has been investigated by performing undrained heating tests while measuring pore pressures changes in a specially adapted thermal isotropic compression cell. Special care was devoted to calibration procedures to account for the effects of the system on the pore pressure measurements. The thermal pressurization coefficient measured appeared to change with temperature, mainly because of the changes with temperature of both the water thermal expansion coefficient of water and the drained compression coefficient of the claystone.

## 1. Introduction

Clays and claystones are considered as potential host rocks for the storage of exothermal high activity radioactive waste at great depth in various countries including France (Callovo-Oxfordian claystone), Belgium (Boom clay) and Switzerland (Opalinus claystone). One of the possible consequences of the temperature elevation caused by the heat emitted in the host rock by the waste is the thermal pressurization of pore water.

Thermal pressurization develops because the thermal expansion of water is much larger than that of the solid phase of the rock. It occurs in low porosity rocks in which heat propagation is much faster than water transfers due to their low permeability ($10^{-20}$ m² in the Callovo-

Oxfordian claystone). Thermal pressurization leads to a decrease in effective stresses that could cause some instabilities related to shear failure or hydraulic fracturing, in particular in the excavation damaged zone (EDZ) close to the galleries.

Further understanding on thermal effects in claystones has been gained from in-situ thermal experiments that have been carried out in Underground Research Laboratories (URL), in particular through the HE-D experiment that was carried out in the Opalinus claystone in the Mont Terri URL in Switzerland [4 5 6] and, more recently, in the TER experiment in Meuse/Haute-Marne URL of Bure in France [3]. In-situ pore pressure measurements indicated that pore water pressure increased from about 1 MPa to 4 MPa when the rock mass was heated up to a temperature of 100°C at a distance of 1 m from the heater. Local rates of pressure increase of 0.16 MPa/°C could be estimated. The TER experiment showed that pore pressure increase rate in the Callovo-Oxfordian claystone could be 1.5 time higher than in the Opalinus claystone [3]. Experimental evidence of thermal pore pressure in the laboratory is scarce, especially for the Callovo-Oxfordian claystone and the determination of relevant parameters is most often made from back analyzing the thermal pore pressures measured during in-situ experiments [6].

In this paper, a special experiment that has been specifically developed for the investigation of the thermal pressurization in the Callovo-Oxfordian (COx) claystone is presented together with the result of the testing program conducted.

## 2. Characteristics of the Callovo-Oxfordian claystone

The ANDRA underground research laboratory of Meuse-Haute-Marne, located near the village of Bure in the North-east of France [7 8] is composed of galleries excavated at a depth of 445 m and 490 m in a 200 m thick subhorizontal (1°-1.5° tilting) layer of the COx claystone, an indurated clay rock dated 155 million years (limit upper-middle Jurassic). This layer is located in between two several hundred meters thick layers of Dogger (bottom) and Oxfordian (top) limestones. The COx claystone is characterised by a very low hydraulic conductivity that restricts water transfer and by a low diffusion coefficient that significantly delays solute transport. The claystone also has a low deformability and its high sorption capacity for radionuclide makes it a proper potential site to store high activity radioactive waste at great depth. The in-situ state of stress at 490m has been investigated in detail [7] and the following stress values have been obtained: vertical total stress $\sigma_v = 12.7$MPa, minor horizontal total stress $\sigma_h = 12.4$MPa and major horizontal total stress $\sigma_H = 12.7$  14.8MPa. In-situ pore pressure measurements provided a value about $u = 4.9$MPa.

The mineralogical composition of the COx claystone depends on the depth with significant changes in carbonate and clay contents. Its total connected porosity varies between 14% in carbonated levels and 19.5% in more argillaceous levels [9]. At 490m, about in the middle of the median sequence of the Callovo-Oxfordian formation (440 – 550m), the claystone contains 40-50% clay (50-70% interstratified illite/smectite), 18-32% quartz, 22-30% carbonate (calcite), less than 2% pyrite, about 5% feldspar, less than 1% of organic materials [10].

Some characteristics of the COx claystone have been provided by Escoffier (2002) [11] who determined a drained isotropic compressibility $C_d = 0.42$GPa$^{-1}$ (drained bulk modulus $K_d = 2$ 410MPa), $C_s = 0.095$GPa$^{-1}$ and a permeability of $2\times10^{-20}$m$^2$ under stress conditions close to in-situ conditions (isotropic unloading from 8 to 6MPa with a back pressure of 1MPa). A value of the linear thermal expansion coefficient of solid grains of $1.4\times10^{-5}$ (°C)$^{-1}$ is given by Gens et al. (2007) [6] corresponding to a volumetric thermal expansion of $4.2\times10^{-5}$ (°C)$^{-1}$.

## 3. Thermal pressurization in clays and claystones

The thermal pressurization of pore water in low porosity clays or claystones submitted to temperature elevation is a consequence of the significant difference between the thermal expansion coefficient of water and that of the solid grains. Ghabezloo and Sulem (2009) [2] gathered some typical values of the thermal expansion coefficients of water and of some typical minerals that are presented in Table 1. The Table is completed by the values of the compressibility of each mineral. The average fractions of each mineral obtained in a sample excavated from the mid-height of the median sequence of the COx formation [10] are also reported in the Table.

One can see that the water expansion coefficient is almost one order of magnitude higher than that of the minerals and that the coefficient of quartz and clay are comparable and larger that that of calcite or feldspar.

Given that the structure of the COx claystone is characterized by a clay matrix containing the grains of quartz, calcite and feldspar [9], one can suspect some significant differential expansions in the solid phase at clay-calcite interfaces (the most frequent) and also at clay-felspar interfaces.

Ghabezloo and Sulem (2009) [2] gathered some values of the thermal pressurization coefficient $\Lambda$ (MPa/°C) measured in various soils and rocks and presented in Table 2. The Table has been completed by their own values on the Rothbach sandstone, by a value on Boom clay recently obtained by Lima et al. (2010) [18] and by a value on Opalinus claystone deduced from the data of Muñoz et al. (2009) [19].

Some large values have been obtained in rocks but the values in clays are between 0.01 and 0.1MPa/°C. Quite different values are given for clays, in particular in Boom clay, comparing the data of Vardoulakis et al. (2002) [20] (obtained from experimental data of Sultan (1997), [25]) and that of Lima et al. (2010) [18]. Authors showed that the thermally induced pore pressure did not only depend on the mineral composition and porosity of the rock, but also on the stress state, the range of temperature variation and the previously induced damage. The pressure dependency of the compressibility of both rock and water and the temperature dependency of the pore water compressibility appeared to play an important role, as shown by Ghabezloo and Sulem (2009) [2] who provided values between 0.25 and 0.025MPa/°C at temperatures between 20 and 70°C for the Rothbach sandstone.

## 4. Material and methods

In-situ, the COx claystone is saturated. However, laboratory samples are desaturated by extraction, storage, transport and laboratory trimming. The mechanical properties of the claystone are highly dependent on the water content with significant increase in unconfined compression stress (UCS) with smaller water content. Pham et al. (2007) [26] performed UCS tests under controlled relative humidity (RH) and obtained UCS values between 27 MPa at 98% RH and 57 MPa at 32% RH. Given its low permeability and swelling properties, the saturation of the COx claystone is a tedious and long process that will be considered in detail in this work, given its importance for a proper determination of the thermal pressurization coefficient.

The two specimens studied here, named EST27396 n°1-iso and EST27396 n°2-iso come from the core EST27396 that has been extracted at a depth of 500 m in the Bure URL. With an initial water content of 6.4% (obtained from weighing the sample before and after a period of 24h in an oven at 105°C) and a porosity of 22% (determined from careful measurements of the sample volume by using a precision calliper), the samples were significantly desaturated with a degree of saturation about 58% (giving a saturated water content of 11%) and a suction of 29MPa (measured by means of a dew point hygrometer, [27]).

The principle of the experiment allowing the determination of the thermal pressurization coefficient in a low porosity claystone is simple: a sample submitted to in-situ stress conditions is submitted to a temperature elevation in undrained conditions while the excess pore pressure is measured by a pressure transducer. However, its completion and the data interpretation appeared to be more complex than anticipated. A schematic view of the device developed for this purpose is described in Figure 1. The system (Figure 1a) is composed of an isotropic compression cell (already used by Tang et al. (2008) [28] to investigate the thermal behaviour of compacted bentonites), connected to two high pressure pressure-volume controllers (PVC, GDS Brand) used to impose the back pressure (maximum pressure 60MPa) and the confining pressure (maximum pressure 60MPa), respectively.

The cell is designed to accommodate a cylindrical sample of 80 mm in diameter and 10 mm in height similar to an oedometric sample. This shape of the specimen has been chosen to minimize the drainage length (equal to the sample thickness, i.e. 10mm) so as to allow satisfactory sample saturation within a reasonable period of time. A short drainage length is also suitable to optimize pore pressure homogeneity and measurements in a very low permeability material. To reduce the parasite volumes that could affect the pore pressure measurements, no porous disk was used and the draining system has been reduced to a simple thin geotextile placed between the bottom base and the sample. The geotextile is connected to two ducts that allow proper saturation, as will be commented later. The drainage performances of the geotextile were checked by using a dummy metal sample under a confining pressure of 20MPa. Also, a specially designed cylindrical neoprene membrane able to continuously envelop the top and lateral face of the sample was used, with no need of using any piston or porous disk on the top of the sample. The membrane is tightly fixed to the bottom base by means of two O-rings.

As seen in Figure 1b, the cell is immersed in a temperature controlled bath. Pore pressure changes in the sample are measured by a pressure transducer (0 to 10MPa range) placed below the bath to avoid any perturbations due to the temperature changes in the bath. The temperature of the bath is measured by means of a thermocouple. There is no displacement measurement in this apparatus.

One of the two ducts arriving at the sample bottom is connected to the pressure transducer whereas the other one is connected to the back-pressure CPV. This CPV was carefully filled by de-aired water.

**Effect of the mechanical and thermal deformation of the system**

The "undrained" condition is achieved by closing the valves of the cell. It is a condition of no change in the fluid mass of the system, i.e. of the pore fluid and of the water in the drainage system. Given the comparable orders of magnitude of the compressibility and of the thermal expansion coefficients of water (from $27\times10^{-5}(°C)^{-1}$ at 25°C to $63\times10^{-5}(°C)^{-1}$ at 80°C under 4MPa [17]), of the solid grains (about $4.2\times10^{-5}(°C)^{-1}$ [6]) and of the metal cell ($5.2\times10^{-5}(°C)^{-1}$ for stainless steel), the volume changes of all components during both the undrained loading phase and the undrained heating phase have to be considered in detail so as to fully understand the various processes occurring during the test. They include the mechanical and thermal volumes changes of the porous specimen (solid and water), of the fluid contained in the drainage system (water in the geotextile, in the ducts machined the metal cell base, in the valves and the pressure transducer) together with the water exchanges between them, that are governed by the specimen low permeability.

The influence of the drainage system on the measurement of the pore pressure response during "undrained compression" in saturated rocks has been examined by Bishop (1976) [29] and Mesri et al. (1976) [30]. The analysis has been extended to thermally induced pore pressures by Ghabezloo and Sulem (2009) [31]. The main results are briefly recalled here.

In a perfect undrained THM test carried out in an elastic porous material [30 32 33], the pore pressure increase is given by the following expression:

$$\Delta u = B\,\Delta\sigma + \Lambda\,\Delta T \qquad (3)$$

where the Skempton coefficient $B$ and the thermal pressurization coefficient $\Lambda$ are defined by the following equations:

$$B = \frac{(C_d - C_s)}{(C_d - C_s) + \phi(C_w - C_s)} \qquad (4)$$

$$\Lambda = \frac{\phi(\alpha_w - \alpha_s)}{(C_d - C_s) + \phi(C_w - C_s)} \qquad (5)$$

where $\phi$ is the porosity, $C_d$ the drained compressibility of the saturated rock, $C_s$ and $C_w$ the compressibilities of the solid phase and of water, respectively.

Accounting for the effects of the drainage system leads to the following corrected expressions [32]:

$$\Delta u_{mes} = B_{mes}\,\Delta\sigma + \Lambda_{mes}\,\Delta T \qquad (6)$$

with:

$$B_{mes} = \frac{(C_d - C_s)}{\phi(C_w - C_s) + (C_d - C_s) + \frac{V_L}{V}(C_w + C_L)} \qquad (7)$$

$$\Lambda_{mes} = \frac{\phi(\alpha_w - \alpha_s) + \frac{V_L}{V}(\alpha_w - \alpha_L)}{\phi(C_w - C_s) + (C_d - C_s) + \frac{V_L}{V}(C_w + C_L)} \qquad (8)$$

in which $V_L$ is the volume of the drainage system, $C_L$ its compressibility and $\alpha_L$ the thermal expansion coefficient of the drainage system. All these parameters have to be determined by running calibration tests.

In other words, the values of the measured parameters have to be corrected as follows [31]:

$$B_{cor} = \frac{1}{\frac{1}{B_{mes}} - \frac{V_L(C_w + C_L)}{V(C_d + C_s)}} \qquad (9)$$

$$\Lambda_{cor} = \frac{\Lambda_{mes}}{1 + \frac{V_L}{\phi V(\alpha_w - \alpha_s)}\big((\alpha_w - \alpha_L) - \Lambda_{mes} V_L(C_w + C_L)\big)} \qquad (10)$$

Equation (9) is the same as that given by Bishop (1976) [29] in the case of isothermal undrained loading.

## 5. Saturation procedure and "undrained" compression test

To avoid any swelling due to uncontrolled hydration, the sample was placed in its initial state (i.e. not fully saturated) on the dry geotextile [34]. Once the system was mounted and the cell filled with water, the sample was isotropically compressed under 8MPa, a value close to the in-situ effective mean stress.

Before injecting the water to the drainage circuits, the vacuum was applied to the all circuits by the vacuum pomp for the few minutes to vacuum all the air trapped in the circuits and geotextile and between the sample and membrane.

The geotextile was then carefully saturated by infiltrating water under a small pressure. To do so, one of the drainage valves of the cell base was connected to the pore water PVC whereas the other one was kept open so as to allow air evacuation from the dry geotextile. This valve was closed once water started to flow out. The pore pressure and confining pressure were then increased simultaneously of the same slope up to near the in-situ stresses (4MPa and 12MPa respectively). At such a high water pressure the air in the unsaturated sample is solved in the water so no air will be trapped between the sample and the membrane.

Figure 2 presents the results of the infiltration phase of water into the samples (EST27396 n°1-iso, EST27396 n°2-iso), showing that most of the water infiltrated during the first four days. A slow and constant infiltration rate is established afterwards. It is difficult to assess whether this flow rate is due to a micro leak or to the mobilisation of the swelling capacity of the claystone. A similar trend has been observed in Boom clay sample [35] and was related to the mobilisation of swelling.

An undrained isothermal test was then carried out on sample EST28396 n°1-iso to determine the Skempton coefficient and to assess the quality of the saturation. To do so, the valves were closed and the confining stress increased at a rate of 0.001MPa/mn while monitoring the changes in pore pressure. Figure 3a shows that the response in pore pressure is nicely coupled to the change in confining stress, a trend confirmed by the diagram of Figure 3b that allows the measurement of the Skempton coefficient $B_{mes} = \Delta u_{mes}/\Delta \sigma_{iso} = 0.7$.

The total volume of the drainage system $V_L$ was directly measured by using the pressure-volume controller. To do so, the drainage system was first dried (by flushing it with a flow of compressed air), put under vacuum and the valves were closed so as to maintain vacuum. The pressure-volume controller and the connecting ducts were filled with de-aired water with no air bubbles trapped in and the duct was carefully connected to the closed valve while setting the volume of the pressure-volume controller at zero. The valve was then gradually opened and the volume of water penetrating the drainage system was given by the pressure-volume controller, giving a value of $V_L$ equal to 2683mm$^3$.

The determination of the $C_L$ coefficient was carried out by conducting a compression test between 4.5 and 7MPa on a dummy metal sample of the same dimensions as that of the specimen. The volume change monitored by the back-pressure PVC while increasing the back pressure depends on the compressibility of both the internal drainage system (inside the cell and limited by the valves) and the external drainage system (the pressure-volume controller itself together with the ducts connected to the cell). To separate these effects, the response of the external system was monitored by running a similar test with the valve closed and the volume change of the internal drainage system $\Delta V_L$ was obtained from the difference between the volume changes obtained from these two tests.

The coefficient $C_L$ is deduced from the expression giving the change in volume of the drainage system with respect to the changes in pore pressure:

$$\frac{dV_L}{V_L} = (C_L + C_w)du \qquad (11)$$

Knowing the value of $C_w$ (0.447GPa$^{-1}$ at 25°C [17]) and the changes in $V_L$ calculated from the response of the pressure volume controller, a value $C_L = 1.6$GPa$^{-1}$ was obtained. This value appears to be significantly larger than the value of 0.117GPa$^{-1}$ obtained by Ghabezloo and Sulem (2009) [31] who used a metal porous stone. This difference is due to the much softer geotextile used here.

The value of $B_{mes}$ was calculated based on the following values of other parameters:
- $C_w = 0.447$GPa$^{-1}$ for water at 25°C [17];

- $C_d = 0.42 \text{GPa}^{-1}$ [11];
- $C_s = 0.095 \text{GPa}^{-1}$ [11];
- $\phi = 0.22$ (measured).

By using Bishop's expression (Equation (9)), a corrected value $B_{cor} = 0.85$ was obtained. This value is considered representative of a good saturation of the COx claystone.

Figure 4a presents the device with the drainage system containing the connecting lines, the valves and the pressure transducer. Figure 4b is a simplified schematic representation of the device (with both the rock sample in a) and the dummy specimen in b)) in which the various volumes are also given.

## 6. "Undrained" heating test

The "undrained" heating was performed by increasing temperature by steps of two degrees and by keeping temperature constant at each step during 10 hours. This period of time is long enough so as to reach temperature equilibrium between the system and the sample.

The specific thermal parameters of the device were determined by performing a thermal calibration test on the metallic dummy sample ($\phi = 0$, $C_d = C_s$). Equation (8) reduces to:

$$\Lambda_{mes} = \frac{\alpha_w - \alpha_L}{C_w + C_L} \qquad (12)$$

in which coefficients $\alpha_w$ and $C_w$ depend on temperature as shown in Figure 5 (after Spang, (2002); [17]).

The result of the dummy thermal test is given in Figure 6. The response shows a reasonably reversible response in temperature. The curve that has been calculated by adopting a value $\alpha_L = 5 \times 10^{-5} (°C)^{-1}$ appears to fit well with the experimental data.

The response in pore pressure with respect to time obtained on a COx sample (EST27396 n°1-iso) submitted to a 2°C step increase in temperature between 25 and 72°C with the valves closed is presented in Figure 7 together with the data of the dummy sample tests already presented in Figure 6. Between 25°C and 42°C, heating was faster with short steps that only lasted 20 minutes whereas between 42 and 72°C steps lasted 10 hours. The data of Figure 8 show that the pore pressure responses obtained during the dummy test are instantaneous and stable at each step, showing good water tightness of the system. Whereas the successive instantaneous pressure increases appear to be comparable in both tests, their cumulated increase with the dummy sample gives a pressure value that becomes clearly larger than that obtained with the COx specimen. The COx specimen response starts with a peak followed by a decrease similar in shape to that of a pore pressure dissipation. The curve also shows that final stabilization is obtained at the end of the 10h long steps.

These trends are observed in more details in the zoom presented in Figure 8. The fast thermal pore pressure increases in both cases are close between 25°C and 42°C with the short 20 mn steps. Conversely, the two curves diverge during the first 10 h step at 42°C with a subsequent decrease in pore pressure in the COx sample.

Various observations can be made:

- The instantaneous thermal pore pressure increments, similar in the dummy and the COx tests, are mainly due to the expansion of the water in the drainage system and of the drainage system;
- The subsequent decrease in the measured pore pressure in the COx test indicates that there is a progressive water transfer from the drainage system into the sample till reaching pore pressure equilibrium between the system and the sample. At equilibrium, the pore pressure gauge gives the value of the sample pore pressure;

- The thermal increase in pore pressure in the sample is coupled with a decrease in effective stress that can be calculated once the Biot coefficient $b$ of the claystone is known. The determination of the $b$ coefficient is not straightforward and a range of values have been provided by various authors. Homand et al. (2006) [36] give values starting from 1 around 10MPa and decreasing to 0.6 around 20MPa. Based on these data, a value of $b = 1$ can be adopted since the initial confining stress and back pressures applied here were initially 12MPa and 4MPa respectively;
- The combined effects of the thermal expansion of water, of the solid grains and the release of effective stress all contribute to the sample expansion that could unfortunately not be directly monitored. This expansion is governed by the drained compressibility $C_d$.

At equilibrium, all thermo-mechanical volume changes are taken into account in Equation (8) that shows that the response is the combined effect of the thermo-mechanical response of the sample solid phase (parameters $C_s$ and $\alpha_s$), of water ($C_w$ and $\alpha_w$), of the drainage system ($C_L$ and $\alpha_L$) and also of the drained compressibility of the claystone ($C_d$).

To complete the data of the previous test in the zone of small temperatures a second test (EST27396 n°2-iso) was conducted with the 2°C temperature steps lasting at least 10 hours along the whole temperature range between 25 and 80°C. The response in pore pressure obtained is presented in Figure 9.

The equilibrated pore pressures at each temperature are representative of the thermally induced pore pressure at the given temperature. The decrease in Terzaghi effective stress corresponding to the increase in pore pressure is presented in Figure 10. It decreases from 8 to 3.4MPa in test EST27396 n°2-iso between 25 and 80°C and from 6 to 4.6MPa in test EST27396 n°1-iso between 43 and 72°C.

The measured thermal pressurisation coefficient ($\Lambda_{mes} = \Delta u_{mes}/\Delta T$, the slope of the pore pressure/temperature curve) has been determined by using two methods: one consisted in taking th derivative of a fitted three order polynomial function whereas the second method (discrete derivative method) was based on a difference quotient calculated on a discrete set of N points centred on the corresponding point.

The corrected thermal pressurization coefficient $\Lambda_{cor}$ was afterwards obtained by applying Equation (10). Conversely to the important effect of the drainage system on the determination of the Skempton coefficient, the drainage system has a small effect on the thermal pressurization coefficient.

The thermal pressurization coefficient decreases with temperature from 0.11 to 0.06 MPa/°C between 42 and 56°C in test EST27396 n°1-iso and from 0.14 to 0.1 MPa/°C between 32 and 62°C in test EST27396 n°2-iso, with respective minimum values observed at 56 and 61°C (Figure 11).

The corrected thermal pressurization coefficients ($\Lambda_{corr}$) obtained from both tests are plotted together in Figure 12 and compared to the theoretical values calculated from Equation (5), in which the changes of parameters $\alpha_w$ and $C_w$ with temperature (Figure 5) are accounted for (with the drained compressibility $C_d$ taken constant and equal to 0.42GPa$^{-1}$ (value at 25°C according to [11]), $C_s$ and $\alpha_s$ respectively equal to 0.095 GPa$^{-1}$ [11], and $4.2 \times 10^{-5}$(°C)$^{-1}$ [6], $\phi = 0.22$ (measured)).

One can observe that the theoretical values do not correspond with the experimental points. The assumption that parameters $C_s$ and $\alpha_s$ are not changing significantly with temperature is commonly admitted (e.g. Gens et al, 2007; [6]). The parameter most prone to change with stress and temperature is the drained compressibility $C_d$. The dependency of $C_d$ with stress is well known, particularly during unloading in materials exhibiting a swelling capacity like the COx claystone, as recently observed in oedometer tests by Mohajerani et al. (2011) [37] who confirmed that $C_d$ significantly increased upon unloading. The change in $C_d$ with temperature is less documented in claystone, but it has been shown by Campanella and Mitchell (1968) [1] in a clay and more recently confirmed by Sultan et al. (2002) [25] in Boom clay that the drained compressibility was not significantly temperature dependent. In the lack of any existing data concerning claystones and in a purpose of simplification, the observation made from clays is adopted here and only the stress dependency of the drained compressibility is considered in a first approach. By fitting the theoretical values of the thermal pressurization coefficient (from Equation (5)) and the experimental values at each temperature, the following expression of the changes in tangent drained compressibility with temperature is obtained.

$$C_d = \frac{\phi(\alpha_w - \alpha_s)}{\Lambda_{cor}} - \phi C_w + (1+\phi)C_s \qquad (13)$$

In Figure 13, one can see that $C_d$ increases from 0.3GPa$^{-1}$ up to 1.1GPa$^{-1}$ during unloading from 8 to 3.4MPa in test EST27396 n°2-iso whereas it increases from 0.4GPa$^{-1}$ up to 1.6GPa$^{-1}$ during unloading from 5.4 to 4.15MPa in test EST27396 n°1-iso. These results are reasonably comparable with those obtained by Escoffier (2002) [11] in isotropic drained compression tests (0.42GPa$^{-1}$ between 8 and 6MPa and 0.65GPa$^{-1}$ between 6 and 4MPa with a 1 MPa back pressure) and those obtained by Bemer et al. (2004) [38] and Mohajerani et al. (2010) [37] from oedometric tests. Note however that the peak observed in the data of test EST27396 n°1-iso does not seem too realistic, given that $C_d$ changes with stress are monotonic.

Using the above values for the drained compressibility $C_d$, the volumetric strain (not monitored here as mentioned above) can be evaluated as follows, admitting the Biot coefficient equal to 1:

$$\varepsilon_v = -\frac{\Delta V}{V} = C_s \Delta u + C_d (\Delta\sigma - \Delta u) - \alpha_s \Delta T \qquad (14)$$

During undrained heating $\Delta\sigma = 0$ and $\Delta u = \Lambda_{cor}\Delta T$, so equation (15) becomes:

$$\varepsilon_v = -\frac{\Delta V}{V} = \alpha_u \Delta T = ((C_s - C_d)\Lambda_{cor} - \alpha_s)\Delta T \qquad (15)$$

The calculated volume change with respect to temperature increase is shown in Figure 14a for both tests. The changes in the undrained thermal expansion coefficient $\alpha_u$ of the claystone with respect to temperature (calculated by taking the derivative of a fitted third degree polynomial expression) are given in Figure 14b. The order of magnitude of the $\alpha_u$ coefficient changes between $7 \times 10^{-5}$ °C$^{-1}$ at 25°C and $14.7 \times 10^{-5}$ °C$^{-1}$ at 80°C. Logically, the $\alpha_u$ values are higher than the average order of magnitude of the thermal expansion coefficient of solid grains ($\alpha_s = 4.2 \times 10^{-5}$ °C$^{-1}$) because of the significant influence of the water thermal expansion ($\alpha_w = 27 \times 10^{-5}$ °C$^{-1}$ at 25°C) and of its changes with temperature between 25 °C and 80°C (at which $\alpha_w = 63 \times 10^{-5}$ °C$^{-1}$, see Figure 5)

## 7. Conclusion

During the deep storage of exothermic radioactive waste, pore fluid pressurization can be induced in the host rock due to the difference between the high thermal expansion of pore water compared to that of the solid phase. In this paper, an experimental evaluation of the thermally induced pore fluid pressurization has been performed on samples of the Callovo-Oxfordian claystone extracted at a depth of 490 m from the ANDRA Meuse-Haute Marne Underground Research Laboratory of Bure. Tests have been conducted in an isotropic compression cell with a sample of small thickness (10mm) so as to reduce the drainage length and optimize both the saturation process and the homogeneity of the pore pressure field. Resaturation of samples was performed under effective stress condition close to the in-situ ones at 490m depth (mean total stress of 12MPa and pore pressure of 4MPa). This technique developed in clayey soils appeared to be necessary so as to avoid any disturbance in the sample due to swelling.

The effects of the mechanical and thermal deformation of the drainage system and of the water that it contains have been considered in the analysis of results of the undrained tests. The correction proposed appeared to be more significant for the evaluation of the mechanical undrained properties than for the thermal ones. It was shown that the thermal pressurization coefficient of COx claystone slightly decreased with increased temperature and with values between 0.15 and 0.1 MPa/°C. This temperature dependency was interpreted as the combined effect of the temperature dependency of the pore water thermal expansion coefficient together with that of the stress dependency of the drained compressibility $C_d$ of the COx claystone. The stress dependency of the $C_d$ parameter was back calculated and favourably compared to values already published.

The data obtained through the completion of a new experiment confirmed that thermal pressurisation depended upon complex interactions including the thermal expansion of the constituents (water + minerals), their change with temperature in the case of water, together with the stress conditions that interact through the stress dependency of the drained compressibility parameter $C_d$. The data obtained here now need to be confirmed and completed by further experimental data related in particular to the temperature dependency of the involved parameters, and more particularly of the drained compressibility.

The numerical simulations of in-situ thermal experiments that have been conducted up to now are based on parameters that are back-calculated to fit with the in-situ measured data. Given the small strain fields involved around the waste during thermal in-situ tests, most calculations are logically carried out in the elastic domain, with constant mechanical parameters. The more detailed insight presented here about the temperature dependency of the different interactive physical phenomena involved in thermal pressurisation will hopefully help to improve the parameter determination and the numerical modelling of the in-situ tests.

## Acknowledgements
The authors wish to acknowledge ANDRA (the French Radioactive Waste Management Agency) for its financial support. The views expressed in this paper are that of the authors and do not engage ANDRA in any matter. The authors also wish to thank Dr. S. Ghabezloo for useful discussions and MM. E. De Laure, H. Delmi and X. Boulay for their invaluable contribution in the development of the new devices used in this work.

*Table 1: Thermal expansion and compressibility coefficients of the COx main minerals.*

| Mineral | Thermal expansion coefficient (°C)$^{-1}$ | Solid compressibility (GPa$^{-1}$) |
|---|---|---|
| Clay | $\alpha_s = 3.4 \times 10^{-5}$ (McTigue 1986) [12] | $C_s = 0.02$ (Skempton 1960 [13], McTigue 1986 [12]) |
| Quartz | $\alpha_s = 3.34 \times 10^{-5}$ (Palciauskas and Domenico, 1982) [14] | $C_s = 0.0265$ (Bass 1995) [15] |
| Calcite | $\alpha_s = 1.38 \times 10^{-5}$ (Fei 1995) [16] | $C_s = 0.0136$ (Bass 1995) [15] |
| Felspar | $\alpha_s = 1.11 \times 10^{-5}$ (Fei 1995) [16] | $C_s = 0.0145$ (Bass 1995) [15] |
| Water | $\alpha_w = 27 \times 10^{-5}$ (Spang 2002) [17] | $C_w = 0.447$ (Spang 2002) [17] |

Table 2: *Thermal pressurisation coefficients of various rocks and soils.*

| Material | Thermal expansion coefficient $\Lambda$ (MPa/°C) | Reference |
|---|---|---|
| Clay | 0.01 | Campanella and Mitchell (1968) [1] |
| Boom Clay | 0.06<br>0.019 | Vardoulakis et al. (2002)[20]<br>Lima et al. (2010) [18] |
| Opalinus claystone | 0.1 | Muñoz et al. (2007) [19] |
| Sandstone | 0.05 | Campanella and Mitchell (1968) [1] |
| Kayenta Sandstone | 0.59 | Palciauskas and Domenico (1982) [15] |
| Rothbach sandstone | From 0.25 to 0.025 | Ghabezloo and Sulem (2009) [2] |
| Clayey fault gouge | 0.1 | Sulem et al. (2004, 2007) [21-22] |
| Intact rock at great depth | 1.5 | Lachenbruch (1980) [23] |
| Mature fault at 7 000m depth | Intact fault wall : 0.92<br>Damaged fault wall : 0.31 | Rice (2006) [24] |

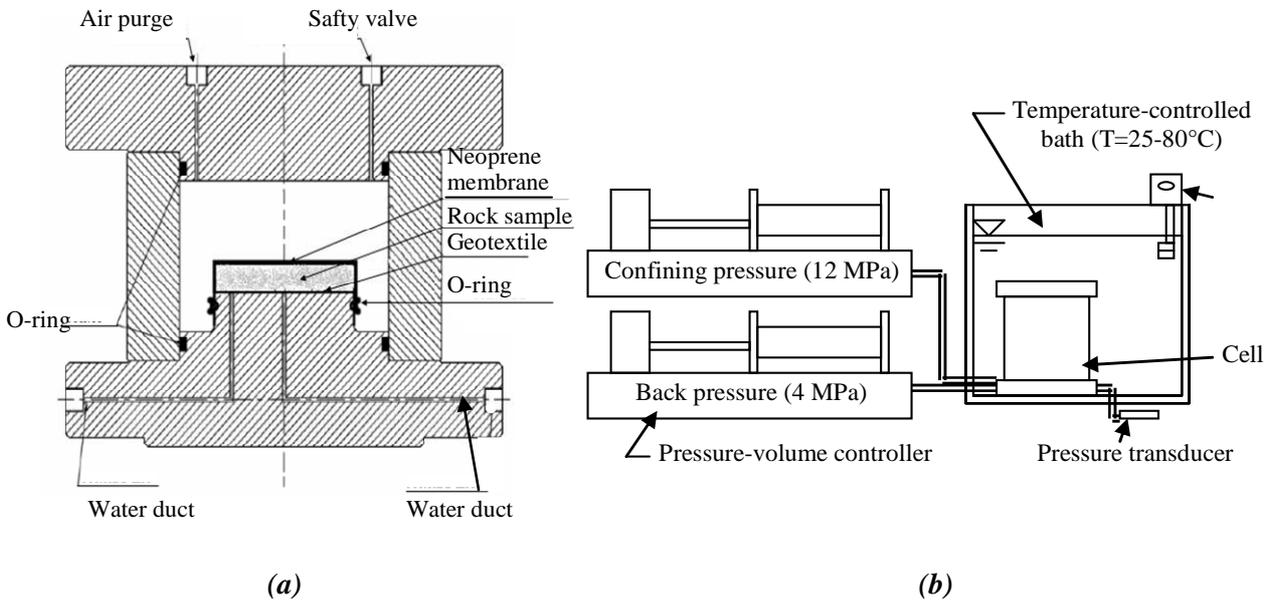

*(a)* *(b)*

Figure 1: *(a) Modified isotropic cell; (b) Controlled temperature bath.*

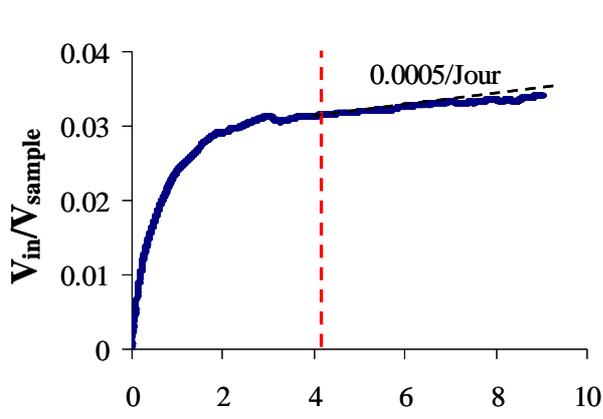 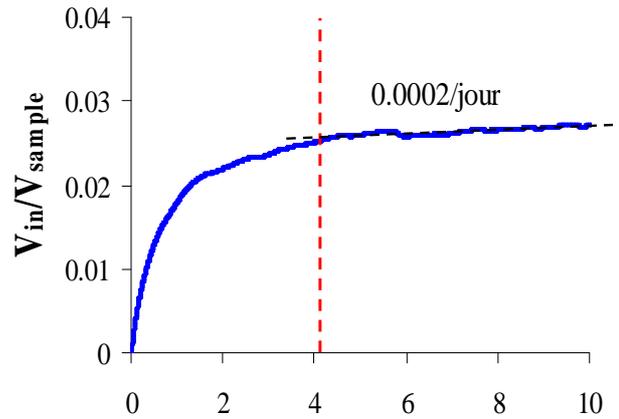

1 *Figure 2: Water injection during the saturation phase: EST28396 n•1-iso and EST28396 n•2-iso.*
2
3
4
5
6

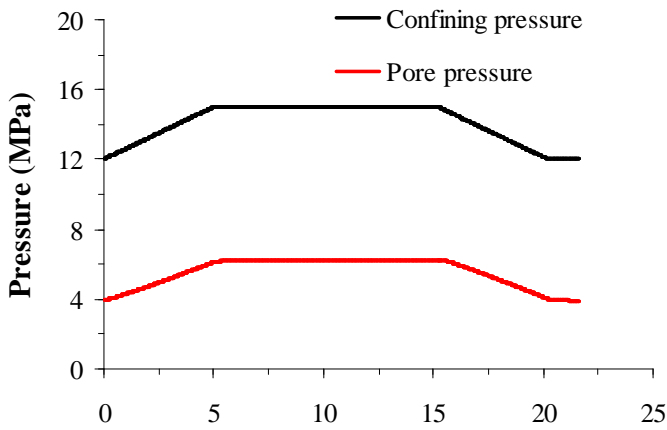 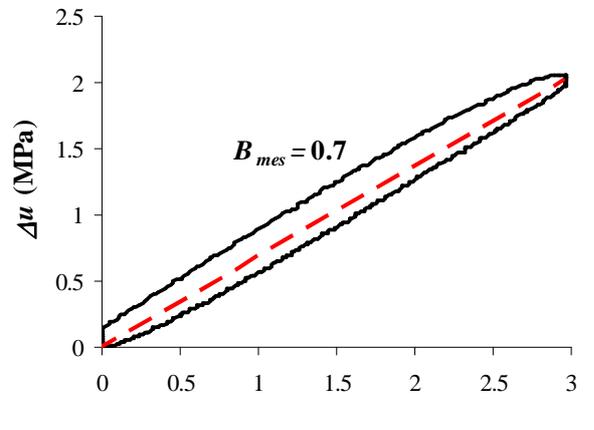

7 *Figure 3: Undrained isotropic compression test: (a) change in pore pressure with time ;(b) change*
8 *in pore pressure under increased confining pressure.*

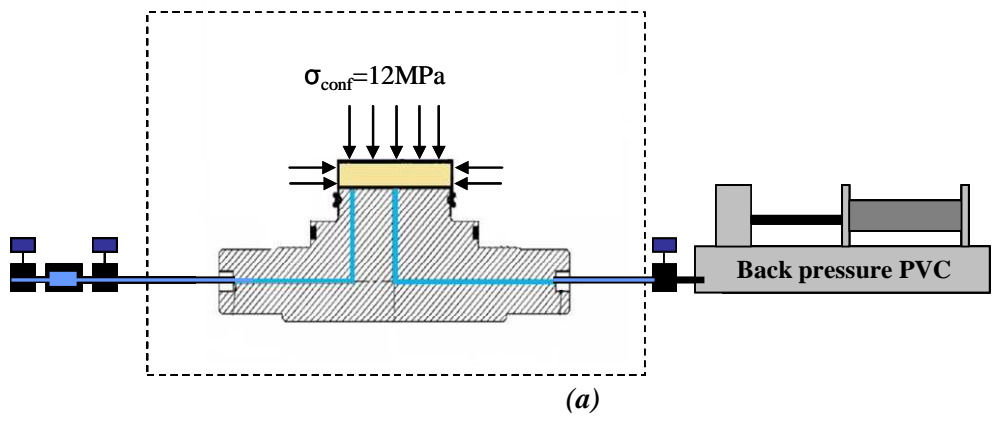

*(a)*

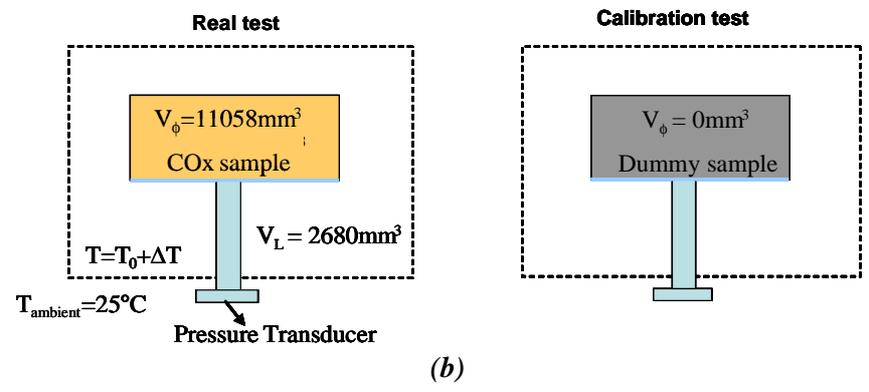

*(b)*

*Figure 4: (a) Representation of the drainage system, (b) Schematic representations of the device*

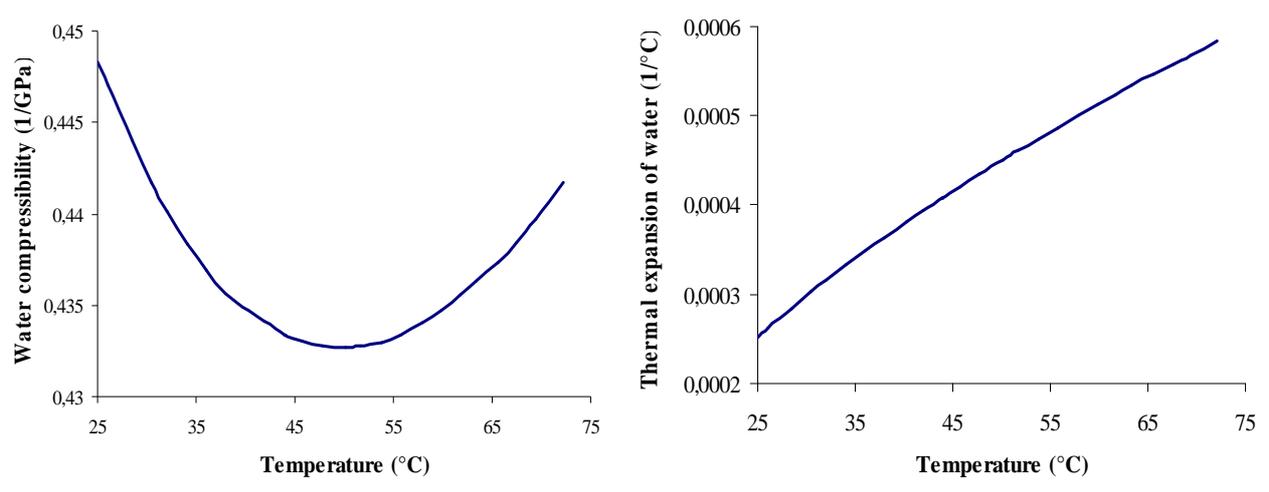

*Figure 5: Compressibility and thermal expansion of water under a 4MPa pressure as function of temperature (after Spang (2002); [17]).*

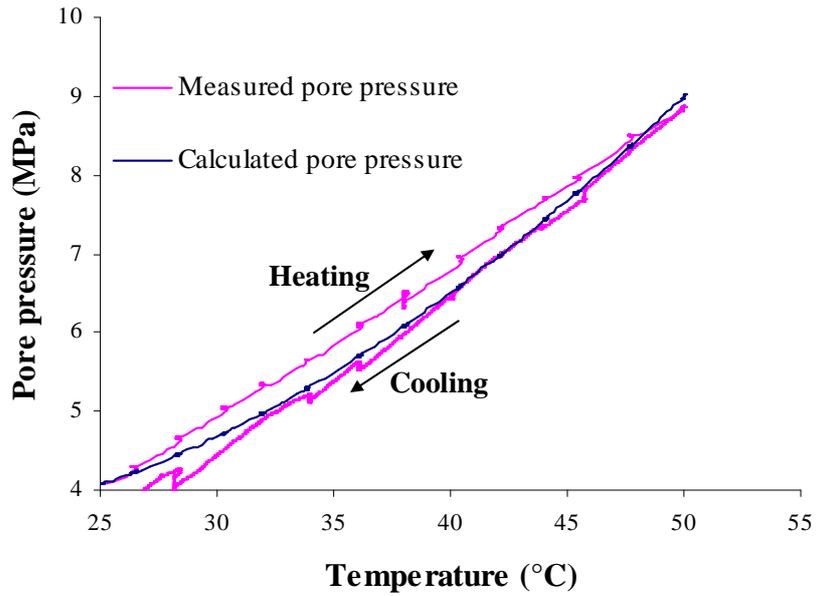

*Figure 6: Thermal pore pressure in the drainage system during a calibration heating-cooling test on a dummy metal sample.*

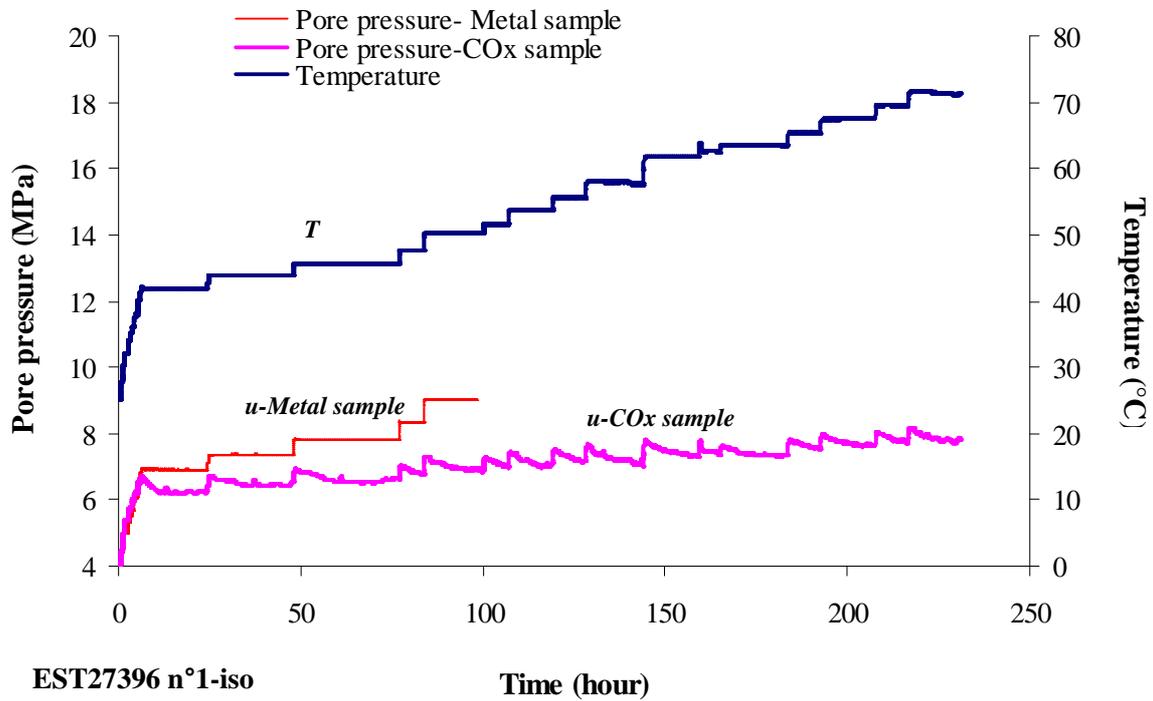

**EST27396 n°1-iso**

*Figure 7: Comparison between thermally induced pore pressure during a test (sample EST27396n•1-iso) and during the calibration test.*

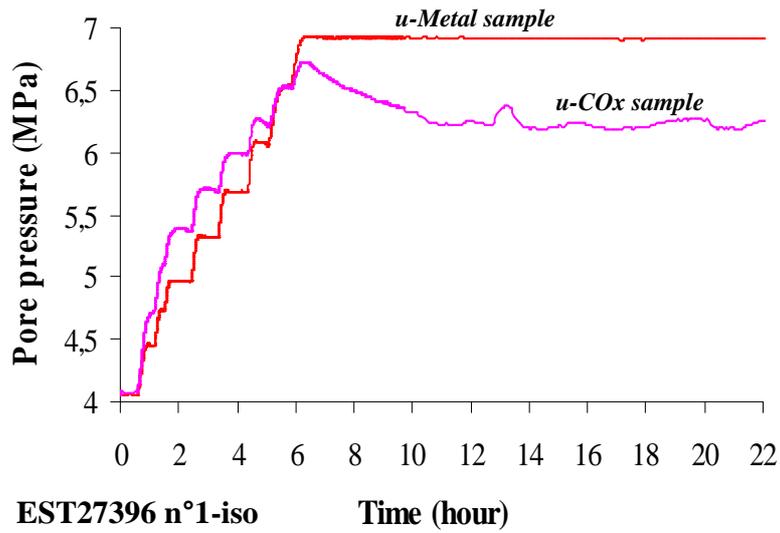

Figure 8: Zoom on the first heating steps from 25°C to 42°C (EST27396 n°1-iso).

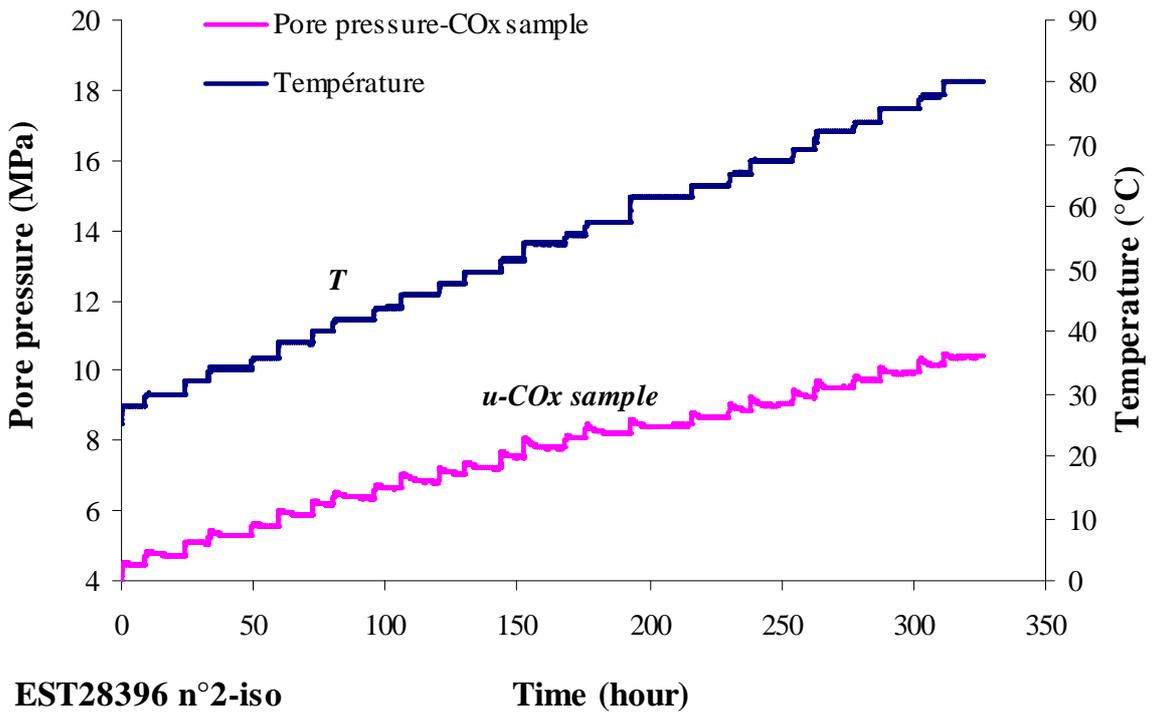

Figure 9: Thermally induced pore pressure in the second sample (EST27396 n°2-iso).

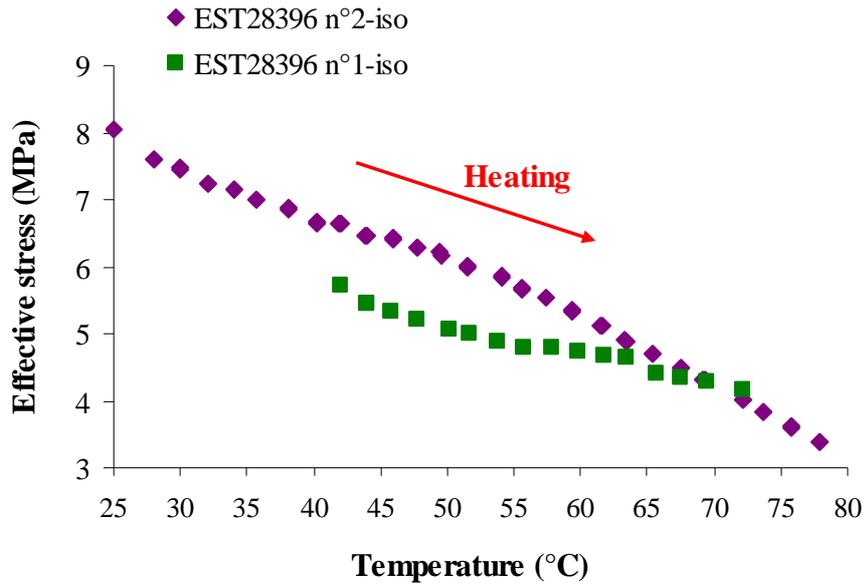

*Figure 10: Decrease in effective stress with temperature.*

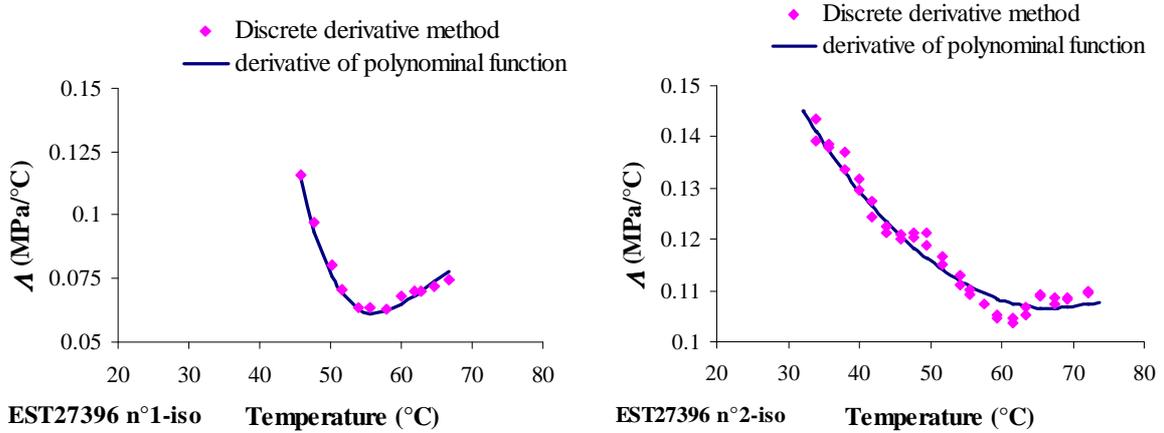

*Figure 11: Calculated values of the changes in thermal pressurization coefficient versus temperature.*

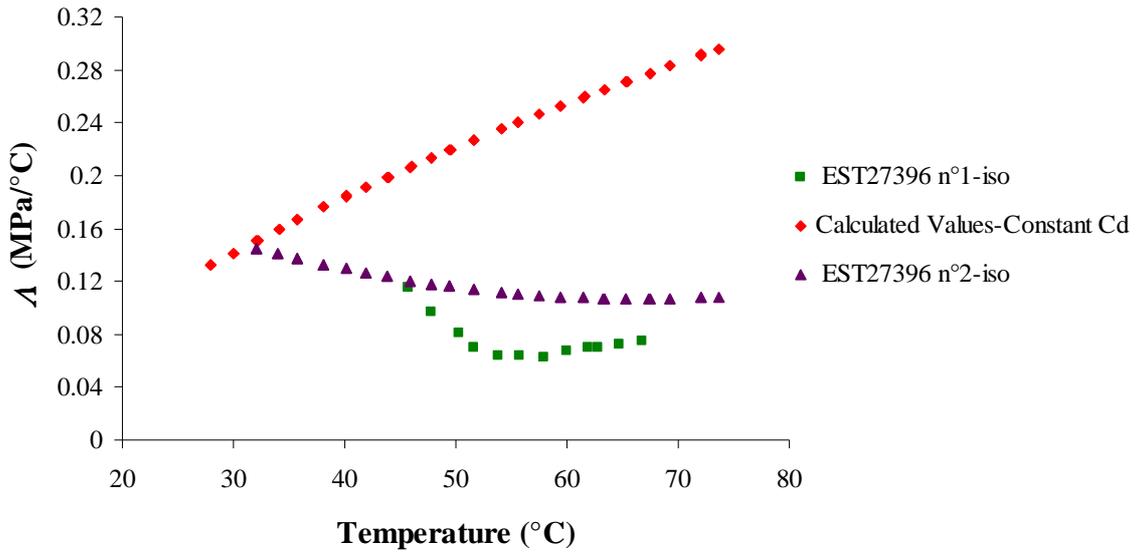

*Figure 12: Thermal pressurization coefficient: experimental values and calculated values, assuming a constant drained compressibility.*

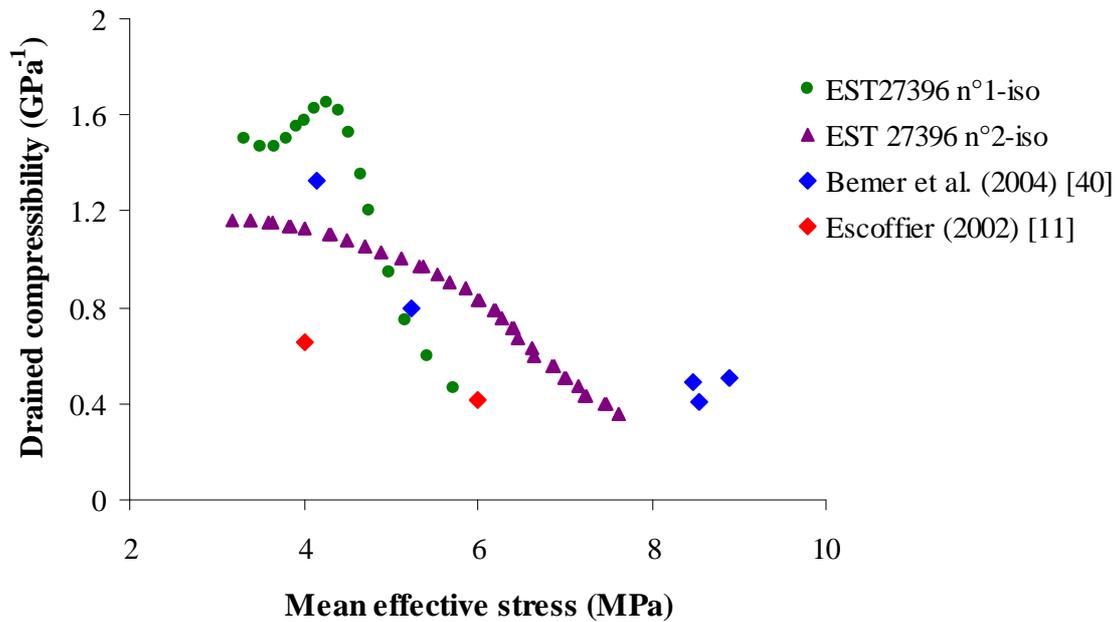

*Figure 13: Changes in drained isotropic compressibility with mean effective stress.*

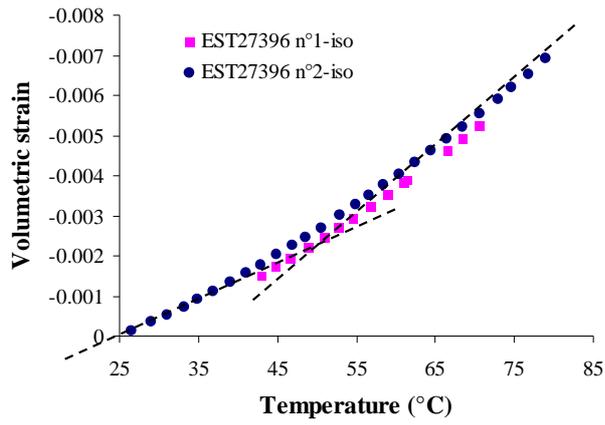 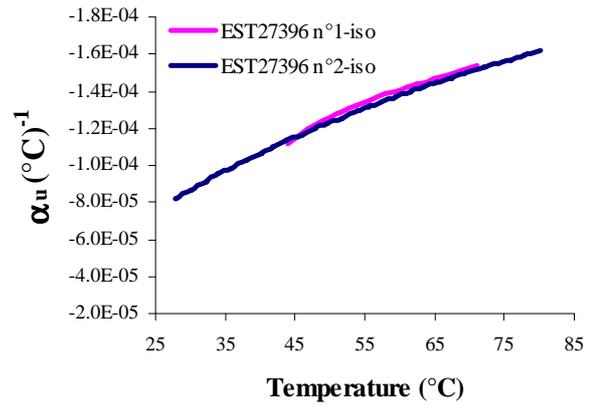

*(a)* *(b)*

*Figure 14: Volumetric strain change during undrained heating tests.(a) Calculated volumetric strain during heating; (b) Undrained thermal expansion coefficient change with temperature.*